\documentclass[twocolumn]{aastex631}

\usepackage{hyperref}
\usepackage{graphicx}	
\usepackage{color}
\usepackage{amsmath}
\usepackage{bm}

\def\apjl{ApJL}

\def\aap{A\&A}

\newcommand{\bdot}{{\displaystyle \cdot}}
\newcommand{\CD}{{\mathcal{D}}}

\newcommand{\CQ}{{\mathcal{Q}}}
\newcommand{\CR}{{\mathcal{R}}}

\newcommand{\QD}{{\mathcal{Q}_\mathcal{D}}}
\newcommand{\RD}{{\left\langle \mathcal{R} \right\rangle_\mathcal{D}}}
\newcommand{\VD}{{\mathcal{V}_\mathcal{D}}}

\newcommand{\avg}[1]{{\left\langle {#1} \right\rangle_\mathcal{D}}}
\newcommand{\avgdot}[1]{{\left\langle {#1} \right\rangle^{\;\bdot}_\mathcal{D}}}
\newcommand{\Rsp}{\mathcal{R}}
\newcommand{\initial}[1]{{#1_{\rm \bf i}}}

\definecolor{mybloodyred}{rgb}{0.7,0,0}
\definecolor{mygreen}{rgb}{0,0.6,0}
\definecolor{myblue}{rgb}{0.2,0.3,0.7}

\begin{document}

\title{Backreaction and the role of spatial curvature in the cosmic neighborhood}

\author[0000-0003-2783-3603]{Marco Galoppo}
\email{Marco.Galoppo@canterbury.ac.nz}
\affiliation{School of Physical \& Chemical Sciences, University of Canterbury, Private Bag 4800, Christchurch 8140, New Zealand}

\author[0000-0002-0828-3901]{Thomas Buchert}
\email{Thomas.Buchert@ens-lyon.fr}
\affiliation{Universit\'e Lyon 1, ENS de Lyon, CNRS, CRAL, UMR 5574, Lyon, France}

\author[0000-0001-8078-6901]{Pierre Mourier}
\email{Pierre.Mourier@canterbury.ac.nz}
\affiliation{School of Physical \& Chemical Sciences, University of Canterbury, Private Bag 4800, Christchurch 8140, New Zealand}
 
\begin{abstract}

We present the first direct computation of spatially averaged dynamical quantities in the local Universe, employing the Cosmicflows-4++ reconstruction and a covariant scalar averaging formalism. We extract the domain-averaged density, expansion rate, spatial curvature, and kinematical backreaction over cosmologically relevant domains around our Galaxy, extending up to a comoving radius of $300~\mathrm{Mpc}/h$. The resulting domain-averaged present-day energy budget features nontrivial variations with scale that reflect a nested structure within the cosmic neighborhood, including a large-scale void shell encompassing the local cosmic web. Remarkably, we find significant contributions to this energy budget from the average spatial curvature at the $\mathcal{O}(10\%)$ level on all probed scales. By contrast, the kinematical backreaction remains much smaller throughout the surveyed volume, reaching at most a $\mathcal{O}(1\%)$ contribution on the smallest scales considered, i.e., $30~\mathrm{Mpc}/h$. Convergence to the global $\Lambda$CDM background is not observed within this range of scales. 

\end{abstract}


\section{Introduction}\label{sec:int}
The late-time Universe features a complex large-scale structure in its matter distribution. This cosmic web can be charted in the region surrounding our Galaxy through the inference of peculiar velocity (PV) flows --- induced by  both baryonic and dark matter --- and direct observations of the baryonic matter density~\citep[see e.g.,][]{York_2000,Erdogdu_2006,Huchra_2012,Courtois_2012,Howlett_2022,Courtois_2025,DESI_2025}. Such cosmographic efforts enable constraints on the predictions of cosmological models for cosmic structure formation and on their parameters~\citep[e.g.,][]{Carrick_2015,Boruah_2020,Said_2020,Courtois_2023,Hollinger_2024}. They also provide a better understanding of our cosmic neighborhood and its impacts on other cosmological probes, such as the Galaxy's peculiar motion contribution to the cosmic microwave background (CMB) dipole~\citep[see][and references therein]{Aluri_2023}. These impacts can be assessed, e.g., by casting the observed surrounding cosmic web structure into a simplified, more tractable model, using exact inhomogeneous and/or anisotropic solutions of the Einstein equations, as well as multipolar expansions of the distance-redshift relation~\citep{Heinesen_2021,Celerier_2024,Giani_2024,Sakr_2024,Koksbang_2025,Kalbouneh_2025,Galoppo_2025,Hills_2026,A.Heinesen_T.Clifton_2026,S.M.Koksbang_2026}. 

The presence of dynamical, spatially inhomogeneous structures in the energy distribution also has broader consequences. Most significantly, in any relativistic cosmological spacetime with fluid sources, inhomogeneities induce backreaction effects on the expansion and evolution of the source flows. Namely, the average dynamics of a given ensemble of fluid elements will not in general match the evolution of an homogeneous-isotropic model universe with the same (initial) average energy densities. This can be seen by describing the effective dynamics of an inhomogeneous fluid within a 3+1 foliation of spacetime through the explicit averaging of local variables over the region of fluid considered within each spatial slice. Such a relativistic averaging formalism and the resulting average dynamics and backreaction terms for a domain comoving with the fluid flow were introduced by~\cite{Buchert_2000,Buchert_2001} for an irrotational dust or perfect-fluid source seen within its rest frames. They were later generalized to arbitrary fluid contents and spatial slices by~\cite{BuchertMourierRoy_2018,BuchertMourierRoy_2020}.

The actual quantitative relevance of these backreaction terms for the large-scale dynamics of our Universe has been the subject of much debate~\citep[see, e.g.,][]{Ellis_1984,Buchert_2008,GreenWald_2011,BenDayan_2013,GreenWald_2014,Buchert_2015,Fleury_2017,CliftonSussman_2019,Adamek_2019,Macpherson_2019,Ginat_2026}. In this Letter, we focus on the low-redshift region of the Universe, where sufficient observational constraints have enabled three-dimensional cosmographic reconstructions of the local velocity field. As such a reconstruction enables the evaluation of kinematics of the cosmic fluid, we can exploit it to estimate the structure-induced backreaction effects within that comoving region of fluid, including the emergence of nonzero spatial curvature. Specifically, we use the latest publicly available Cosmicflows-$4$++ (CF4++) reconstruction~\citep{Courtois_2025} of the present-day PV and density fields up to a comoving distance of $300$ Mpc/$h$, to estimate the local matter flow expansion and shear, as well as the local spatial intrinsic curvature, within that region. This allows us to evaluate the averaged energy budget --- including the contributions from backreaction terms --- on the present-day slice across a range of scales around the observer, using the irrotational dust relativistic averaging framework of~\cite{Buchert_2000}.

We summarize the setup of this averaging scheme and its main features relevant for this work in Sec.~\ref{sec:averaging}.
In Sec.~\ref{sec:reconstructions}, we briefly outline the contents and methodology of the CF4++ reconstruction used in this work as local data. We then present our strategy for mapping these Newtonian fields into a consistent relativistic framework, in particular for evaluating the spatial scalar intrinsic curvature.
In Sec.~\ref{sec:results}, we further detail how we extract the relevant local variables and the contributions to the averaged energy budget (including backreaction terms) at various scales, and we present and discuss the results. Finally, in Sec.~\ref{sec:concl}, we summarize these findings and examine strategies for future improvements. In Appendices~\ref{app:A} and~\ref{app:B} we investigate the robustness of our results against some limitations of the CF4++ reconstruction and of the numerical scheme employed in our analyses, respectively.

\section{Covariant Averaging Scheme}
\label{sec:averaging}

In this Letter, we are interested in structures in the matter-dominated Universe at scales large enough that effects of velocity dispersion and vorticity can be neglected. Hence, we model the Universe in this era as sourced by a single, irrotational dust fluid, and characterize its average kinematic and dynamical properties over given regions of space using the covariant relativistic spatial-averaging scheme for scalars introduced for such a source by~\cite{Buchert_2000}.
We summarize below the construction of this scheme and the key quantities that emerge in the averaged dynamics in the presence of local inhomogeneities.

\subsection{Defining Spatial Averages for Scalars}

We consider a four-dimensional, globally hyperbolic Lorentzian spacetime with metric $\bm{g} = g_{\mu\nu} \, \mathrm{d}x^\mu \mathrm{d}x^\nu$. The energy-momentum tensor $T^{\mu \nu}$  is associated with a single pressureless (dust) fluid source with energy density $\rho$ and irrotational timelike 4-velocity $\bm u = u^\mu \;\! \bm \partial_\mu$, i.e., $T^{\mu\nu} = \rho \, u^\mu u^\nu$.
The metric then defines a local projector orthogonal to the 4-velocity, ${\bm h}$, with components $h_{\mu \nu} := g_{\mu \nu} + u_\mu u_\nu$. The 4-velocity field $\bm u$ is further characterized by its expansion scalar $\Theta := h^{\mu \nu} \nabla_\mu u_\nu = \nabla_\mu u^\mu$, and by its symmetric-traceless shear tensor components $\sigma_{\mu \nu} := h^\kappa_{\,(\mu} {h}^\lambda_{\,\nu)}\nabla_\kappa u_\lambda - (\Theta/3) \, h_{\mu \nu}$ and associated shear scalar $\sigma^2 := (1/2) \, \sigma^{\mu \nu} \sigma_{\mu\nu}$.

By Frobenius' theorem, the irrotational velocity field $\bm u$ defines a foliation of spacetime into spacelike hypersurfaces orthogonal to $\bm u$, representing global rest frames for the dust. Since the 4-velocity thus coincides with the normal to these hypersurfaces, the projector $h$ also corresponds to the spatial metric on these hypersurfaces. We then adopt comoving coordinates $(x^\mu)=(t,X^i)$ adapted to this foliation, where $t$ is the fluid proper time. We denote with an overdot ${}^\bdot$ the local evolution operator along $\bm u$, $u^\mu \nabla_\mu$, and, consistently, the time derivative of any scalar variable $\varphi(t)$ that depends only on time, $\dot\varphi := \mathrm{d}\varphi(t) / \mathrm{d}t$.

Note that the averaging scheme remains covariantly defined through the above characterization of the coordinate set. Generalizations to other fluid sources and foliations, as well as manifestly covariant expressions, of this formalism can be found in~\cite{BuchertMourierRoy_2018,BuchertMourierRoy_2020,Gasperini_2009,Gasperini_2010,HeinesenMourierBuchert_2019,Fanizza_2020}.

An arbitrary compact domain $\CD$ within the spatial slices can then be selected as the region of interest over which average properties are computed. The propagation of the domain between slices is defined to be \emph{comoving with the fluid flow}, so that it encompasses a fixed set of fluid elements throughout their evolution.

The volume of the averaging domain $\CD$ is defined as its Riemannian volume from the induced spatial metric on the hypersurfaces:
\begin{equation}
   \label{eq:defVD}
    \VD(t) := \int_\CD \, \sqrt{h} \; \mathrm{d}^3X \, , \;  h := \det(h_{ij}) \, , \; i,j = 1,2,3 \, .
\end{equation}

One can then define the corresponding volume-weighted spatial average $\avg{\psi}$ of any scalar variable $\psi$ over the domain of interest:
\begin{equation}
    \avg{\psi}(t) := \frac{1}{\VD(t)} \int_\CD \; \sqrt{h} \; \psi \; \mathrm{d}^3 X \; .
\end{equation}

The fluid-comoving propagation of the domain implies that the evolution rate of its volume is directly given by the averaged expansion scalar: $\dot{\mathcal{V}}_\CD / \VD = \avg{\Theta}$. More generally, from this assumption the following evolution equation can be derived for the average of any scalar variable $\psi$, here expressed as a \emph{commutation rule} between the spatial averaging and time-evolution operators:
\begin{equation}
    \label{eq:comm_rule}
    \left\langle \vphantom{\dot\psi} \psi \right\rangle_\CD^{\displaystyle .} - \avg{\dot\psi} = \avg{ \vphantom{\dot\psi} \Theta \, \psi} - \avg{\vphantom{\dot\psi} \Theta} \, \avg{ \vphantom{\dot\psi} \psi} \; .
\end{equation}

\subsection{Averaged Evolution Equations}

From the above definition of the volume of the averaging domain, an \emph{effective scale factor} $a_\CD$ can be defined at all times, tracking the evolution of the characteristic size of this comoving domain: $a_\CD(t) \propto \VD(t)^{1/3}$. The evolution rate of this scale factor defines an effective Hubble function $H_\CD(t)$ that depends on $\CD$:
\begin{equation}
    H_\CD(t) := \frac{\dot a_\CD}{a_\CD} = \frac{1}{3} \frac{\dot{\mathcal{V}}_\CD}{\VD} = \frac{1}{3} \avg{\Theta} \; .
\end{equation}

The dynamical evolution equations for the effective scale factor are obtained by spatially averaging the energy constraint, the Raychaudhuri equation, and the dust density conservation equation. Using the commutation rule~\eqref{eq:comm_rule}, one obtains, respectively,
\begin{align}
   \label{eq:avg_Hamilton}
    & 3 \frac{\dot a_\CD^2}{a_\CD^2} =  8 \pi G \avg{\rho} + \Lambda - \frac{1}{2} \RD - \frac{1}{2} \QD \; ; \\
   \label{eq:avg_Raych}
    & 3 \frac{\ddot a_\CD}{a_\CD} = - 4 \pi G \avg{\rho} + \Lambda + \QD \; ; \\
    \label{eq:avg_energyconserv}
    & \avgdot{\rho} + 3 \frac{\dot a_\CD}{a_\CD} \avg{\rho} = 0 \; .
\end{align}
The (independent) last equation follows from the conservation of the total fluid rest mass within the domain comoving with the fluid flow, more directly expressed as $\avg{\rho} \VD = const$.

Compared to the Friedmann equations, these evolution equations feature an additional term, the \emph{kinematical backreaction}, $\QD$, which measures the inhomogeneity and anisotropy of the fluid's volume deformation:
\begin{equation}
   \label{eq:defQD}
    \QD := \frac{2}{3} \left( \, \avg{\Theta^2} - \big\langle{\Theta} \big\rangle^{\,2}_\CD \, \right) - 2 \, \avg{\sigma^2} \; .
\end{equation}
Additionally, the energy balance equation~\eqref{eq:avg_Hamilton} depends on another nontrivial term: the average of the local scalar intrinsic 3-curvature $\Rsp$ on the hypersurfaces. The average curvature does not in general scale as $\propto a_\CD^{-2}$; rather, its evolution is coupled to structure formation via the kinematical backreaction, obeying the conservation law or \emph{integrability condition},
\begin{equation}
  \label{eq:intcond1}
    a_\CD^{-6} \left(a_\CD^6 \, {\mathcal{Q}}_\CD  \right)^\bdot+ a_\CD^{-2} \left( a_\CD^2 \, \RD \right)^\bdot = 0 \; ,
\end{equation}
which is obtained by combining the time derivative of Eq.~\eqref{eq:avg_Hamilton} with Eqs.~\eqref{eq:avg_Raych} and \eqref{eq:avg_energyconserv}. Jointly, the two coupled contributions $\QD$ and $\RD$ encompass the deviations of the effective scale factor's evolution from that of the corresponding homogeneous Friedmann--Lema\^itre-Robertson--Walker model --- specifically, a flat $\Lambda$--cold dark matter ($\Lambda$CDM) universe model. Hence, with respect to such a spatially flat reference or background model, both terms are to be considered as backreaction effects impacting the regional dynamics.

The above system (Eqs.~\eqref{eq:avg_Hamilton}--\eqref{eq:avg_energyconserv} and \eqref{eq:intcond1}) may be viewed as a set of balance equations that quantify the contributions and dynamical coupling of curvature and inhomogeneities, either globally or over a given averaging scale. It is therefore useful to rewrite the averaged energy constraint~\eqref{eq:avg_Hamilton} on a domain $\CD$ as a sum of dimensionless $\Omega^\CD(t)$ functions that depend on the averaging domain and scale, by normalizing the equation by $3 H_\CD^2(t)$:
\begin{equation}
     \Omega^\CD_M + \Omega^\CD_\Lambda + \Omega^\CD_\mathcal{R} + \Omega^\CD_\mathcal{Q} = 1 \, ,
 \label{eq:sumOmegas}
\end{equation}
where we have defined
\begin{align}
     &\Omega^\CD_M := \frac{8 \pi G \;\! \big\langle \rho \big\rangle_\CD}{3 H_\CD^2}  \, ; \quad \Omega^\CD_\Lambda := \frac{\Lambda}{3 H_\CD^2} \, ; \\
     &\, \Omega^\CD_\mathcal{R} := - \frac{\RD}{6 H_\CD^2} \, ; \quad \Omega^\CD_\mathcal{Q} := - \frac{\QD}{6 H_\CD^2} \, . \label{eq:omegas} 
\end{align}
The present-day value (at time $t=t_0$) for the domain $\CD$ of a given $\Omega^\CD_i(t)$ is denoted as $\Omega_i^{\CD,0}$.

\section{Local Universe Modeling}
\label{sec:reconstructions}

We model the local Universe using the reconstructed present-day density and PV fields from the CF4++ catalog presented by~\cite{Courtois_2025}. In the following, we first summarize the key features of this reconstruction. We then outline the correspondence between Newtonian and general-relativistic (GR) models which we employ to extract the spatial averages introduced in Sec.~\ref{sec:averaging}.

\subsection{The CF4++ reconstruction}
\label{subsec:CF4++}
The CF4++ catalog~\citep{Courtois_2025}, expanding the original CF4 dataset~\citep{Tully_2023}, constitutes the largest sample of galaxy PVs to date, providing near all-sky coverage up to a measured redshift $z \simeq 0.05$ (corresponding to a background comoving distance of $\sim 150~\mathrm{Mpc}/h$), and extending up to a cutoff in measured redshift at $z = 0.1$, i.e., $\sim 300~\mathrm{Mpc}/h$, with some data points reaching larger measured luminosity distances of up to $\sim 500~\mathrm{Mpc}/h$.

The CF4++ reconstruction of~\cite{Courtois_2025} provides a direct estimate for the present-day ($t=t_0$) density and PV fields, constrained by the CMB power spectrum and a present-day slice projection of the observed PV data from our past light cone. Crucially, the reconstruction assumes that Eulerian-frame Newtonian linear perturbation theory about a global flat $\Lambda$CDM background provides an accurate description at the scales probed. This assumption allows one to directly relate, in a simplified way, the local density contrast, $\delta$, and the inferred PV field, $\vec{v}_\mathrm{pec}$, via
\begin{equation}
\vec{\nabla} \cdot \vec{v}_\mathrm{pec} + a H f \, \delta = 0 \, , \label{eq:linear_theory}
\end{equation}
at linear order, where $a(t)$ is the background scale factor, $H(t)$ is the corresponding Hubble rate, and $f(t) = \mathrm{d} (\ln \delta) / \mathrm{d}(\ln a)$ is the growth rate. 

Accordingly, the CF4++ reconstruction is presented as a present-day snapshot on a regular comoving grid with a side length of $1~{\rm Gpc}/h$, centered on the observer and divided into $128^3$ nodes, corresponding to a spatial resolution of about $8~{\rm Mpc}/h$ . The (spatially flat) background cosmological parameters for the reconstruction are set to $\Omega_M^0 = 0.3$, $\Omega_\Lambda^0 = 1 - \Omega_M^0$, and $H_0 =74.6~\mathrm{km} {\cdot} \mathrm{s}^{-1} {\cdot} \mathrm{Mpc}^{-1}$. 

In this work, owing both to the redshift cutoff of the data at $z = 0.1$ and to measurement uncertainties increasing rapidly with distance, we restrict our analysis to within a comoving radius of $300~\mathrm{Mpc}/h$, as done in the bulk flow analysis by~\cite{Courtois_2025}. This choice therefore restricts our analysis to a volume in which the reconstruction itself (i) remains constrained by the available data, and (ii) is only weakly affected by the reconstruction requirement that field averages over the whole $\pm 500~\mathrm{Mpc}/h$ box match the $\Lambda$CDM prior.


\subsection{Bridging the Gap between Newtonian and general-relativistic averages}
\label{subsec:mapping}

We aim to recover a relativistic interpretation of the relevant fields extracted from the Newtonian reconstruction of CF4++, and to subsequently derive from them the various components of the averaged energy balance (Eqs.~\eqref{eq:sumOmegas}--\eqref{eq:omegas}) over given matter-comoving regions in the local Universe. As detailed below, this entails little practical difference in the computations compared to fully Newtonian evaluations; however, a well-defined correspondence between Newtonian and GR frameworks is necessary for the formal geometric interpretation of the spatial curvature term. The GR framework also broadens the scope of the kinematical backreaction term, which reduces to a boundary term in the Newtonian case~\citep{BuchertEhlers_1997}.

More specifically, the correspondence strategy employed here has been developed and applied in a series of papers beginning with~\cite{BuchertRZA_2012}, and was subsequently summarized by~\cite{Buchert_2023}. This correspondence was formulated within the same setting adopted here, namely, a flow-orthogonal foliation of spacetime with an irrotational dust matter model. Briefly, this strategy yields an algebraic correspondence between Newtonian self-gravitating fluid dynamics in the Lagrangian frame and the $1+3$ GR equations for dust in fluid-comoving (Lagrangian) spatial coordinates.

The key aspect of this formal correspondence for the purposes of this work is the definition of a local variable corresponding to the fluid rest-frame spatial scalar intrinsic curvature $\Rsp$ in GR, via a direct equivalent to the energy constraint:
\begin{equation}
\label{eq:newtoncurvature}
\Rsp = 2\Lambda + 16\pi G \varrho + \Theta^i_{\ j}\Theta^j_{\ i} - \Theta^2 \, ,
\end{equation}
where $\Theta_{ij}$ denotes the components of the expansion tensor of the velocity field. In GR, these are defined as $\Theta_{ij} = h^\mu_{\,(i} h^\nu_{\,j)} \nabla_\mu u_\nu$. Within the GR--Newtonian correspondence, their counterparts (still denoted $\Theta_{ij}$) are the components in the Lagrangian basis of the Newtonian fluid-velocity field expansion tensor, $\Theta^{N}_{ij}=\partial_{(i} u_{j)}$. Here, since we set the Lagrangian coordinates to match the Eulerian ones on the present-day slice, $\Theta_{ij}$ simply reduces to $ \Theta^{N}_{ij}$ on that slice. Then, in both GR and Newtonian frameworks, the expansion tensor components can be split as $\Theta_{ij} = (\Theta/3) \, h_{ij} + \sigma_{ij}$, where, in the Newtonian case, $h_{ij}$ is set to $\delta_{ij}$, and the expansion scalar is $\Theta = \partial_i u^i$. Here, since the GR spatial metric --- unlike its spatial derivatives --- remains everywhere close to Euclidean on the scales considered~\citep{BuchertEllisvanHelst_2009}, we indeed expect that a Newtonian evaluation of $\Theta_{ij}$ faithfully traces its GR counterpart.

Evaluating the right-hand side of the above equation~\eqref{eq:newtoncurvature} within Newtonian theory generally results in a nonzero left-hand side, but without a direct geometric interpretation, since the Newtonian framework has a vanishing spatial curvature. However, following the GR correspondence established above, this effective, dynamical ``Newtonian curvature'' provides a good estimate of --- and can be interpreted as --- the (flow-orthogonal) spatial scalar curvature $\Rsp$ within a GR setting. Here, we have also incorporated the cosmological constant $\Lambda$ in the Newtonian picture as a straightforward extension, thereby allowing, e.g., the flat $\Lambda$CDM background to be included as an exact solution, as in GR.

Furthermore, since the kinematical backreaction variable $\QD$ has the same expression in terms of expansion rate variance and shear in both theories, we can consistently estimate the GR variable from the corresponding Newtonian fields. 

Additionally, the Newtonian counterpart of the averaged Raychaudhuri equation \eqref{eq:avg_Raych}, as well as the integrability condition \eqref{eq:intcond1} between $\QD$ and the averaged value of the scalar ``curvature" $\Rsp$ in the Newtonian framework, maintain the same form as in GR. Formally integrating the latter equation over time then yields an integral relation between the kinematical backreaction and the averaged $\Rsp$ which also holds in both GR and Newtonian representations~\citep[see][Sec.~2.3.1]{Buchert_2008}:
\begin{equation}
\label{eq:curvature_intcond}
\avg{\Rsp} = \frac{6 \:\! k_\initial{\CD}}{a_{\CD}^2} - \QD - \frac{2}{a_\CD^2} \int_{\initial t}^{t} {\mathrm d}t' \, \QD \frac{\mathrm d}{{\mathrm d}t'} 
a^2_\CD (t') \ ,
\end{equation}
with a domain-dependent integration constant $k_\initial{\CD}$ specified at some initial time $\initial t$. Interestingly, Eq.~\eqref{eq:curvature_intcond} shows that the averaged scalar curvature encodes the full past history of the kinematical backreaction, which explains why the two coupled terms $\QD$ and $\RD$ can have very different magnitudes at a given time. Indeed, quantitative models suggest that curvature will typically dominate over $\QD$, with some estimates indicating a contribution of order unity from the spatial curvature even for small metric perturbations~\citep{BuchertEllisvanHelst_2009}. We can therefore expect the data to exhibit a stronger contribution from $\RD$ than from $\QD$.

\medskip

\section{Results}
\label{sec:results}

We are now in a position to employ our model of the local Universe, together with the correspondence between Newtonian and GR descriptions in the scalar averaging framework, to extract spatial averages of dynamically relevant quantities in our cosmic neighborhood.

We can directly extract the dust density field $\rho$ from the CF4++ reconstruction, given the knowledge of the background cosmological parameters and the reconstructed density contrast field. To obtain the expansion scalar and shear tensor, we first define the Newtonian cosmological velocity field $\vec{u} = H_0\;\!\vec{x}+\vec{v}_\mathrm{pec}$ on the reconstruction grid, which is provided in comoving coordinates $x^i$ on the present-day slice. The expansion scalar and shear tensor components are then computed from the spatial gradients of $\vec{u}$ using a second-order finite-difference scheme, according to
\begin{align}
    & \Theta = \vec{\nabla}\cdot\vec{u} = 3 H_0 + \vec\nabla \cdot \vec v_\mathrm{pec}\; ;  \label{eq:rec_theta}\\
    & \sigma_{ij} =\partial_{(i}u_{j)} -\frac{1}{3}\:\!\!\left(\vec{\nabla}\cdot\vec{u}\right)\delta_{ij} =\partial_{(i} v_{j)}^\mathrm{pec} -\frac{1}{3}\:\!\!\left(\vec{\nabla}\cdot\vec{v}_\mathrm{pec} \right)\delta_{ij} \, . \label{eq:rec_shear}
\end{align}
The corresponding shear scalar is computed as in Sec.~\ref{sec:averaging}, via $\sigma^2 = (1/2) \, \sigma^{ij} \sigma_{ij}$, where indices are raised trivially using the Euclidean spatial metric. We then employ the above Newtonian kinematic and matter variables to evaluate the right-hand side of Eq.~\eqref{eq:newtoncurvature}, determining the spatial scalar curvature $\Rsp$. 

\begin{figure*}[htb!]
    \centering
    \includegraphics[width=\textwidth]{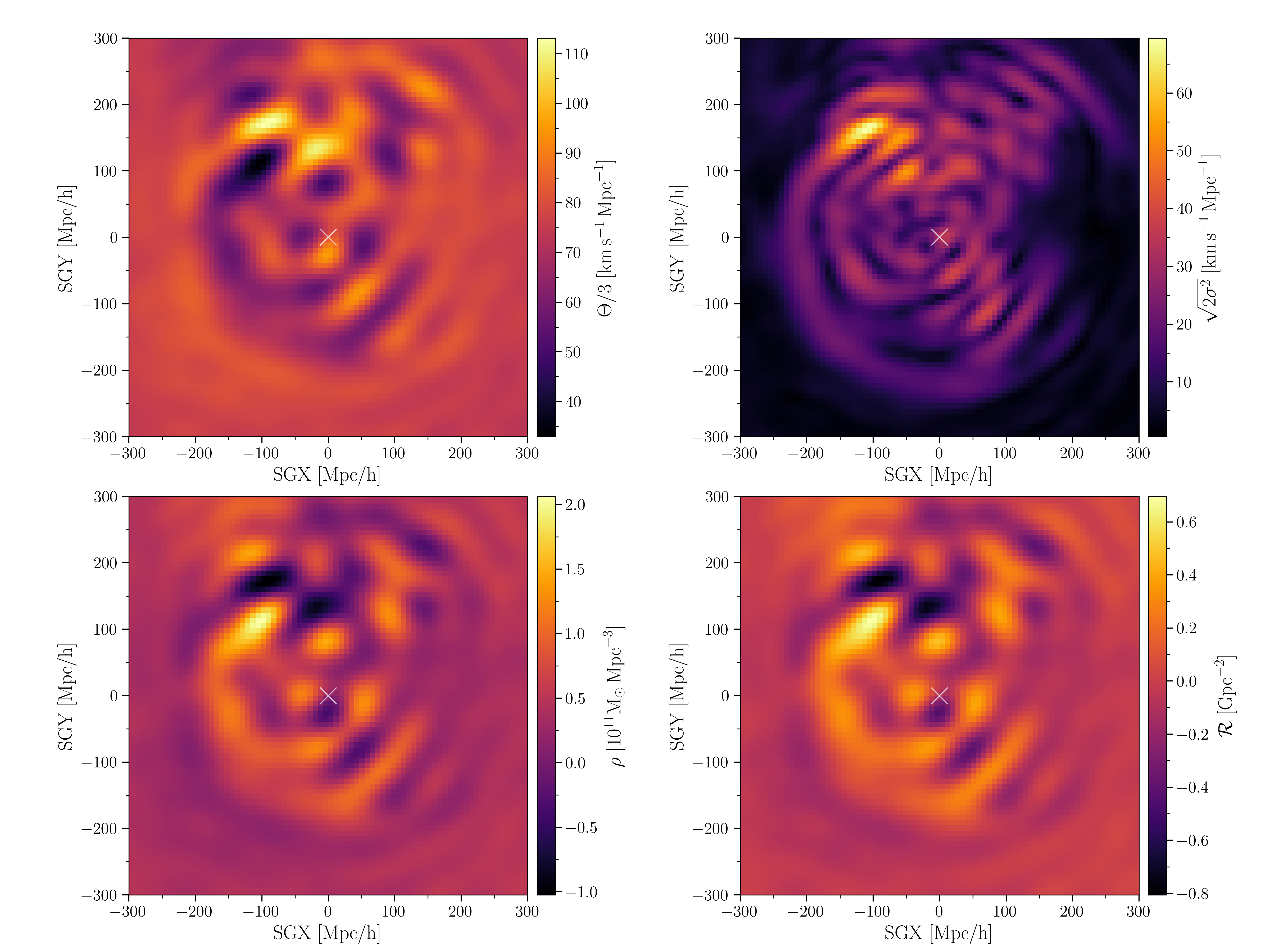}
    \caption{Expansion rate ($\Theta$/3, top-left panel), shear scalar amplitude ($\sqrt{2\sigma^2}$, top-right panel), matter density ($\rho$, bottom-left panel), and spatial scalar curvature ($\Rsp$, bottom-right panel) on the Supergalactic plane as functions of the Supergalactic comoving-distance coordinates SGX, SGY, from the CF4++ reconstruction. In all panels, the white cross marker at the center indicates the observer's position.} 
    \label{fig:AllVariables_FR}
\end{figure*}

As an example, Fig.~\ref{fig:AllVariables_FR} shows two-dimensional slices (corresponding to the Supergalactic plane) of the reconstructed length-expansion rate $\Theta/3$,  shear amplitude $\sqrt{2\sigma^2}$, matter density $\rho$, and the corresponding spatial scalar curvature $\Rsp$, out to a comoving distance of $300~\mathrm{Mpc}/h$. We adopt $\Theta/3$ and $\sqrt{2\sigma^2}$ to facilitate a direct magnitude comparison with the homogeneous expansion rate $H_0$ of the assumed global background.

The maps show that large-scale coherent structures in the local Universe imprint expected correlated signatures across the density, expansion, and curvature fields. Regions of enhanced matter density are systematically associated with reduced expansion, while underdense regions correspond to locally enhanced expansion. Furthermore, the inferred spatial scalar curvature field closely traces the large-scale matter distribution, being negative for underdensities and positive for overdense regions. By contrast, the shear magnitude instead peaks in the surroundings and interfaces of large structures, producing shell- or ridge-like features.

Additionally, within the adopted reconstruction scheme, and at the given grid coarse-graining scale of about $8~\mathrm{Mpc}/h$, the inferred flow is expanding ($\Theta>0$) throughout the surveyed volume. Hence, the reconstruction does not identify any region that has locally fully decoupled from the background expansion ($\Theta\leq0$) at such scales. Although in accordance with estimates placing the size of the largest virialized structures in the Universe at an order of $2~\mathrm{Mpc}/h$~\citep[see, e.g.,][for a review]{Walker_2019}, it more surprisingly also points to the nondetection of significantly larger collapsing regions.

Finally, we note the presence of some unphysical negative-density regions in deep voids (in about $3 \%$ of cells within $300$ Mpc/$h$), as well as some large positive density contrast values within overdense structures that lie outside the linear regime and may accordingly be biased (in about $4 \%$ of cells within $300$ Mpc/$h$). These are artifacts of the linear density reconstruction scheme employed. We assess their impact on our results in Appendix~\ref{app:A}, and show that they do not qualitatively affect our conclusions. 

We now extract the relevant spatial averages for given averaging domains on the present-day slice according to the framework of Sec.~\ref{sec:averaging}, approximating the spatial metric on that slice as Euclidean ($\sqrt{h} \sim 1$ in the comoving coordinates $x^i$). This is thus equivalent to using the Newtonian counterpart of the averaging operator~\citep{BuchertEhlers_1997}. Such an approximation remains fully consistent, since the exact frame defining the spatial slice, whether strictly fluid-orthogonal or featuring small (nonrelativistic) PVs as a tilt, has very little quantitative impact on the evaluation of comoving-domain averages of quantities defined from the matter flow --- such as $\CR$, $\rho$, $\Theta$, or $\sigma^2$~\citep[see][Sec.~4.4]{Mourier_2024}.

\begin{figure*}[htb!]
    \centering
    \includegraphics[width=0.8\textwidth]{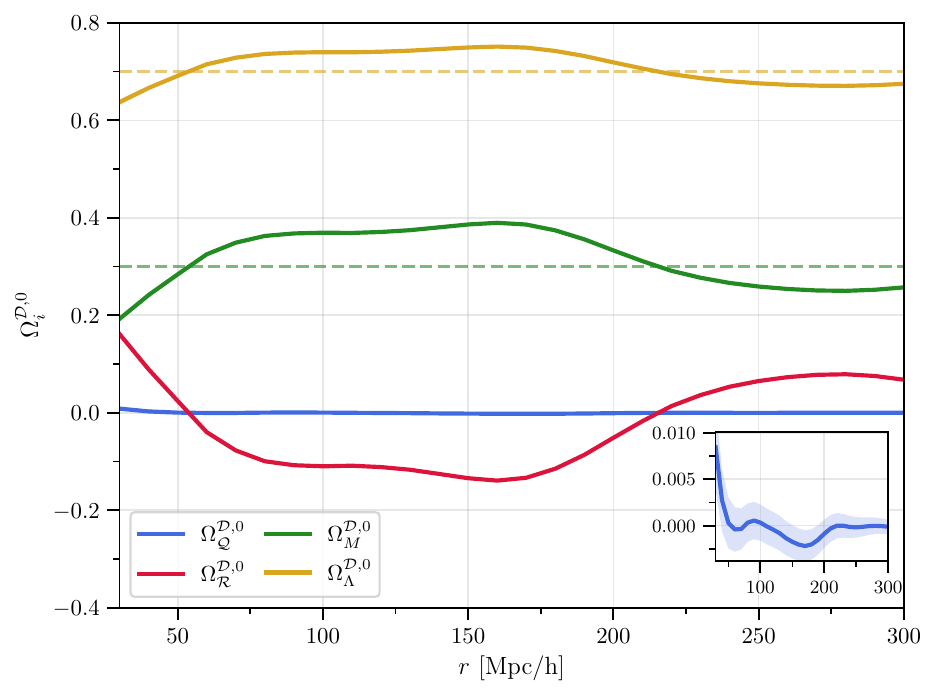}
    \caption{The present-day averaged energy budget as measured by the $\Omega^{\CD,0}_i$ parameters, and computed within the CF4++ reconstruction for concentric spherical averaging domains $\CD$ of varying comoving radius $r$, from $r = 30$ Mpc/$h$ to $r = 300$ Mpc/$h$. The green and yellow horizontal dashed lines represent the $\Lambda$CDM background values used in the reconstruction for $\Omega_M^0$ and $\Omega_\Lambda^0$, respectively. A nonnegligible $\Omega^{\CD,0}_\mathcal{R}$ is observed throughout the surveyed volume, whilst $\Omega^{\CD,0}_\mathcal{Q}$ remains very small with only minor fluctuations (shown in the zoomed-in inset, where the shaded area denotes the $1\sigma$ statistical uncertainties). Convergence to the global $\Lambda$CDM background is not observed within this range of scales.}
    \label{fig:OmegaDi_FR}
\end{figure*}

We consider a range of Lagrangian (comoving with the dust flow) averaging domains $\mathcal{D}$, constructed on the present-day slice as observer-centered spheres with increasing comoving radius, from $30$ to $300~\mathrm{Mpc}/h$, thus probing the dynamics of the corresponding region of fluid. Indeed, the quantitative impact on the local dynamics of the coherent large-scale structures observed in Fig.~\ref{fig:AllVariables_FR}, encoded in spatial averages, is made explicit in Fig.~\ref{fig:OmegaDi_FR}, which shows the present-day values $\Omega^{\CD,0}_i$ of the domain-dependent $\Omega_i^\CD$ parameters (see Eq.~\eqref{eq:omegas}) as functions of the comoving averaging domain radius. Statistical uncertainties on the $\Omega^{\CD,0}_i$ were estimated from the voxel-level error budget provided with the CF4++ reconstruction~\citep{Courtois_2025}, both through direct error propagation and through a Markov Chain Monte Carlo procedure with $10{,}000$ realizations, in which the reconstructed fields are sampled voxel by voxel from the CF4++ mean values and $1\sigma$ deviations. We find these statistical uncertainties to be at the $\delta \Omega^{\CD,0}_i  = \mathcal{O}\!\left(10^{-3}\right)$ level, reflecting the strong suppression of stochastic fluctuations by the spatial-averaging procedure. Consequently, in Fig.~\ref{fig:OmegaDi_FR} they are not visually discernible, except in the zoomed-in inset showing $\Omega_{\mathcal{Q}}^{\CD,0}$. Additionally, we assessed the error on these variables introduced by computing the local kinematical quantities (as PV gradients) on a discrete grid, finding it to be negligible (see Appendix~\ref{app:B}).

Remarkably, across the full range of scales probed, the averaged spatial curvature contribution to the energy budget $\Omega^{\CD, 0}_\Rsp$ remains significant and does not simply converge toward zero with increasing domain size. Rather, we find a nontrivial behavior, with a varying sign and a substantial amplitude up to $300~\mathrm{Mpc}/h$. This suggests that the local Universe does not approach the homogeneous and spatially flat $\Lambda$CDM background within the surveyed volume. Rather, we find that the spatial scalar curvature should contribute at the $\mathcal{O}(10\%)$ level to the present-day effective dynamics even on the largest scales constrained by the reconstruction data, despite an initially nearly vanishing value at the recombination era. This can be understood by noting that the spatially averaged curvature over a domain comoving with the fluid does not obey a separate conservation law, unlike the Friedmannian curvature term (for which $\Rsp \propto a^{-2}$). Instead, its evolution is governed by the coupling with the corresponding $\QD$ in Eq.~\eqref{eq:intcond1}. Consequently, it is not required to ``average out'' --- or rapidly decay with time --- on finite domains. We emphasize that within our framework this remains true, since conditions such as $\left\langle{\Rsp}\right\rangle_\mathrm{box}=0$ (and ${\Omega}^{\mathrm{box},0}_M = 1 - {\Omega}^{\mathrm{box},0}_\Lambda =0.3$) are enforced by the reconstruction scheme, but only on the full $\pm 500~\mathrm{Mpc}/h$ reconstruction box and need not hold for sub-box averaging domains.

Furthermore, we find that the local Universe is, on average, negatively curved ($\Omega^{\CD,0}_{\Rsp}>0$) in our vicinity out to about $50~\mathrm{Mpc}/h$, then transitions to an extended positively curved region ($\Omega^{\CD,0}_{\Rsp}<0$) out to over $200~\mathrm{Mpc}/h$, and becomes again negatively curved on larger scales up to at least $300$ Mpc/$h$. This scale-dependent behavior is consistent with a nested cosmic environment, with a small local void embedded within a large local overdensity, itself enclosed by an even larger underdense shell. This picture appears to be broadly compatible with the presence of observed and/or proposed very large cosmic structures in the local Universe, such as the Sloan Great Wall~\citep{Gott_III_2005,Einasto_2011}, the Ho'oleilana ring ~\citep{R.B.Tully_2023Hola}, and the Local Hole~\citep{R.C.Keenan_etal_2013,Wong_2022}.

However, we emphasize that the large-scale behavior must still be interpreted with care. Indeed, the reconstruction is based on PV data with effective full-sky coverage only out to about $150~\mathrm{Mpc}/h$, while at larger distances the sampling becomes increasingly sparse and anisotropic, and the associated uncertainties grow. Consequently, conclusions about the sign change and amplitude of $\Omega^{\CD,0}_\Rsp$ on the largest scales should be regarded as indicative, given the coverage of the current data.

Finally, we find that the kinematical backreaction contribution $\Omega^{\CD,0}_\mathcal{Q}$ remains subdominant throughout the range of averaging domains surveyed, reaching at most about $1\%$ of the energy budget on the smallest averaging scale considered around the observer ($30$ Mpc/$h$ in comoving radius), as shown in the zoomed-in inset of Fig.~\ref{fig:OmegaDi_FR}. This confirms that, within the present CF4++ reconstruction and at its given spatial resolution, the direct backreaction effects associated with the shear and fluctuations in the expansion of the matter flow remain relevant only on very local scales. Instead, it is their integrated history encoded in the averaged curvature term $\RD$ (see Eq.~\eqref{eq:curvature_intcond}) that dominates nontrivial large-scale dynamics and as such is the major driver of backreaction effects. However, 
since $\QD$ is extracted from a linear reconstruction, its inferred magnitude is expected to be reduced relative to that obtained within a fully nonlinear framework. Therefore, the current results should be understood as a lower limit on the amount of kinematical backreaction in our cosmic neighborhood.

\section{Discussion and Conclusions}
\label{sec:concl}

In this Letter, we presented the first direct measurement of spatially averaged dynamical quantities over varying scales in the local Universe by combining the Newtonian Cosmicflows-$4$++ (CF4++; \cite{Courtois_2025}) reconstruction with a GR scalar averaging framework for irrotational dust~\citep{Buchert_2000,Buchert_2001}. From the CF4++ reconstructed present-day density and peculiar velocity fields, we inferred the local expansion scalar $\Theta$ and shear tensor components $\sigma_{ij}$, and --- by employing the Newtonian--GR kinematic correspondence for self-gravitating systems~\citep{Buchert_2023} --- we further reconstructed the local scalar spatial intrinsic curvature $\mathcal{R}$. This illustrates how large-scale coherent structures imprint correlated signatures across these dynamical fields.

We then evaluated the spatial averages of these fields over concentric spherical domains around the observer, on the present-day slice, with comoving radii ranging from $30~\mathrm{Mpc}/h$ to $300~\mathrm{Mpc}/h$. From the averaged shear scalar and the expansion variance, we also extracted the kinematical backreaction $\QD$ on each domain (see Eq.~\eqref{eq:defQD}). Finally, we expressed the domain-dependent present-day averaged energy budget in terms of the dimensionless $\Omega^{\CD,0}_i$ parameters, and quantified their (small) statistical uncertainties using the CF4++ error model. In particular, we found that the energy contribution from the spatial curvature, $\Omega^{\CD,0}_\Rsp$, is nonnegligible across the full range of scales probed. Indeed, it remains at the $\mathcal{O}(10\%)$ level --- in broad agreement with nonperturbative order-of-magnitude estimates~\citep{BuchertEllisvanHelst_2009} --- with a nontrivial scale dependence in its magnitude and sign. By contrast, the kinematical backreaction contribution $\Omega^{\CD,0}_\mathcal{Q}$ is subdominant throughout the surveyed volume, essentially fluctuating around zero, and reaching at most the $\mathcal{O}(1\%)$ level on the smallest probed scales. 

These values could a priori be compared with estimates from relativistic cosmological simulations~\citep{E.Bentivegna_M.Bruni_2016,J.T.Giblin_etal_2016,Macpherson_2019,C.Tian_etal_2021,A.Oestreicher_S.M.Koksbang_2024}. However, a direct comparison is not possible with the currently available simulation results as (i) the fluid's kinematical backreaction, as defined in Eq.~\eqref{eq:defQD}, has been computed  only by~\cite{E.Bentivegna_M.Bruni_2016}, exclusively on the full simulation domain; and (ii) the intrinsic curvature, when evaluated, was not defined from the fluid frame as in Eq.~\eqref{eq:newtoncurvature}, but rather on specific simulation slices~\citep{Macpherson_2019,C.Tian_etal_2021,A.Oestreicher_S.M.Koksbang_2024}.

Another key result of our analysis pertains to the scale dependence of the averaged energy budget. Remarkably, this shows no convergence toward the imposed, global, spatially flat $\Lambda$CDM background within distances of $300~\mathrm{Mpc}/h$, in agreement with related results from analyses of the bulk flow~\citep[see][and references therein]{Watkins_2023,Courtois_2025} and of galaxy catalogs~\citep{Sylos_Labini_2026}. As mentioned, $\Omega_{\Rsp}^{\CD,0}$ remains nonnegligible and distinctly scale dependent within this range. In parallel, the averaged mass density contribution $\Omega_{M}^{\CD,0}$, and to a lesser extent the fractional contribution from the cosmological constant $\Omega_{\Lambda}^{\CD,0}$, also exhibit appreciable (and related) fluctuations about their respective $\Lambda$CDM background values. This supports the conclusion that the local Universe does not reach statistical homogeneity within the probed volume. This has direct implications for local cosmography, as analyses of low-redshift observables that assume a scale-independent, CMB-constrained, spatially flat $\Lambda$CDM background within the local volume may incur systematic biases. Indeed, strong regional fluctuations in the average dynamics and spatial curvature can modify the effective redshift--distance relation at a nonnegligible level, thereby impacting local cosmological inference.

Additionally, our analysis suggests the presence of a nested structure in the local environment. The varying sign of $\Omega_{\Rsp}^{\CD,0}$, as well as the correlated fluctuations in $\Omega_{M}^{\CD,0}$ around its background value, point to the presence of a local void out to about $ 50~\mathrm{Mpc}/h$, followed by a network of local overdensities (walls) up to $ \sim 200~\mathrm{Mpc}/h$, and, interestingly, the emergence of a large-scale void shell beyond, surrounding the local cosmic web and broadly compatible with the proposed Local Hole. More generally, this illustrates that a scale-dependent analysis of spatially averaged dynamical quantities can serve as a quantitative diagnostic, potentially enabling the direct identification and characterization of coherent cosmic structures.

However, we emphasize that while the overall amplitude of the fluctuations and qualitative scale dependence of the $\Omega_{i}^{\CD,0}$ parameters are robust within our framework (see also Appendix~\ref{app:A}), the finer quantitative details --- e.g., the exact amplitude of $\Omega_{\Rsp}^{\CD,0}$, the precise radii of its sign changes, and its behavior on the largest scales --- remain limited by the current data and by the CF4++ reconstruction methodology.

In particular, the specific assumed background model remains a potential source of systematic bias in the quantitative evaluation of averaged fields. For example, a flat $\Lambda$CDM background forces the globally averaged curvature to be zero. Conversely, any cosmology in which the \emph{global} energy budget itself already includes nonzero contributions from spatial curvature and/or kinematical backreaction (e.g., \cite{Wiltshire_2007,Wiltshire_2009b,HeinesenBuchert_2020,Giani_2025a}) would generically shift the reconstructed regional amplitudes of these terms. In such scenarios, intrinsic curvature and kinematical backreaction partially (or fully) replace the role played by the cosmological constant in driving late-time acceleration. One then typically expects the inferred contributions of $\QD$ and $\RD$ to be correspondingly enhanced relative to the $\Lambda$CDM-calibrated reconstruction used here.

Furthermore, the reconstruction scheme of CF4++ is based on a Newtonian Eulerian-frame linearized framework, which becomes unreliable in the nonlinear regime, systematically misplacing emerging structures and biasing their amplitudes~\citep[see, e.g.,][and references therein]{Melott_1994,MelottBuchert_1994,MelottBuchert_1995}. These limitations call for moving beyond Eulerian linear schemes and toward Lagrangian-frame structure-formation reconstructions. The latter would more reliably capture the large-scale distribution at present day, while ensuring positivity of the reconstructed density field~\citep{Nusser_1991,Gramann_1993,SusperregiBuchert_1997,DoumlerI_2013,DoumlerII_2013,DoumlerIII_2013}. Crucially, such Lagrangian approaches, having already been developed at higher order in Lagrangian deformations within the Newtonian framework, admit fully GR extensions~\citep{BuchertRZA_2012,BuchertRZA_2013,AllesRZA_2015,AlRoumiRZA_2017,LiRZA_2018,GasparBuchert_2021,Buchert_2022,GasparBuchertOstrowski_2023}. This provides a natural route toward improved, self-consistent relativistic reconstructions of the local Universe.

Finally, in the coming years the release of novel high-volume PV data from new surveys, such as DESI and 4MOST~\citep{Taylor_2023}, will substantially improve the depth, sky coverage, and sampling density of PV constraints~\citep[see already, e.g.,][]{Ross_2025}. This will enable reconstructions at higher resolution and out to larger distances. Therefore, future work developing and applying relativistic Lagrangian reconstruction schemes to new PV data should enable a sharper localization of the scale-dependent features in the regional $\Omega_{i}^{\CD,0}$ parameters, a more reliable assessment of convergence toward statistical homogeneity, and a correspondingly more quantitative estimate of the impact of local structure on cosmographic inference. \\

\section*{Acknowledgements}

The authors are grateful to Hélène Courtois, Yehuda Hoffman, and Aurélien Valade for insightful discussions on the methodology and implementation of current reconstruction algorithms. We also thank Leonardo Giani, Asta Heinesen, Sofie Koksbang, Jenny Wagner, and David Wiltshire for insightful comments on the manuscript, and Christopher Harvey-Hawes, Emma Johnson, and Zachary Lane for useful discussions. MG and PM acknowledge support from the Marsden Fund grant M1271 administered by the Royal Society of New Zealand, Te~Ap\=arangi.  This work forms a spin-out of a project that has received funding from the European Research Council (ERC) under the European Union's Horizon 2020 research and innovation program (grant agreement ERC advanced grant 740021--ARTHUS, PI: TB), supporting TB and PM.

\newpage
\bibliography{main}{}
\bibliographystyle{aasjournal}

\appendix

\section{Robustness against Eulerian linear reconstruction artifacts}
\label{app:A}

In this appendix, we assess the impact of known shortcomings in the present CF4++ reconstruction on our results --- most notably the breakdown of Eulerian-frame Newtonian linear theory in the nonlinear regime (i.e., $|\delta| \gtrsim  1$), which leads to unphysical (negative) density values at some places within voids and, more generally, can bias the amplitudes of emerging structures. We perform a set of controlled robustness tests aimed at mitigating such problematic features in the reconstructed density field and assessing their impact on the derived kinematic and curvature quantities.

Specifically, we consider three complementary strategies: (i) a ``quenching'' procedure in which we clip the density contrast such that $\delta\in \left[-1, 1\right]$; (ii) a hard excision in which, when computing spatial averages of any field,  we consistently remove from the integration (and volume computation) all grid cells whose density contrast is in a fully nonlinear regime ($|\delta|>1$); and (iii) a coarse-graining of the underlying density and velocity fields via a sliding-window average over $16 \times 16 \times 16$ grid cells, corresponding to an effective smoothing scale of $125~\mathrm{Mpc}/h$. Within a factor of $2$, this is the smallest window size that ensures the resulting coarse-grained density contrast field remains everywhere within the linear regime, $|\delta| < 1$ (and consequently $\rho > 0$). For each of the three modified-field constructions, we recompute the full set of derived scalars (expansion rate, shear amplitude, and spatial curvature) and repeat the domain-averaging analysis, extracting again the corresponding domain-dependent averaged energy budget parameters $\Omega_i^{\CD,0}$ for the same family of concentric spheres.

\begin{figure*}[htb!]
    \centering
    \fig{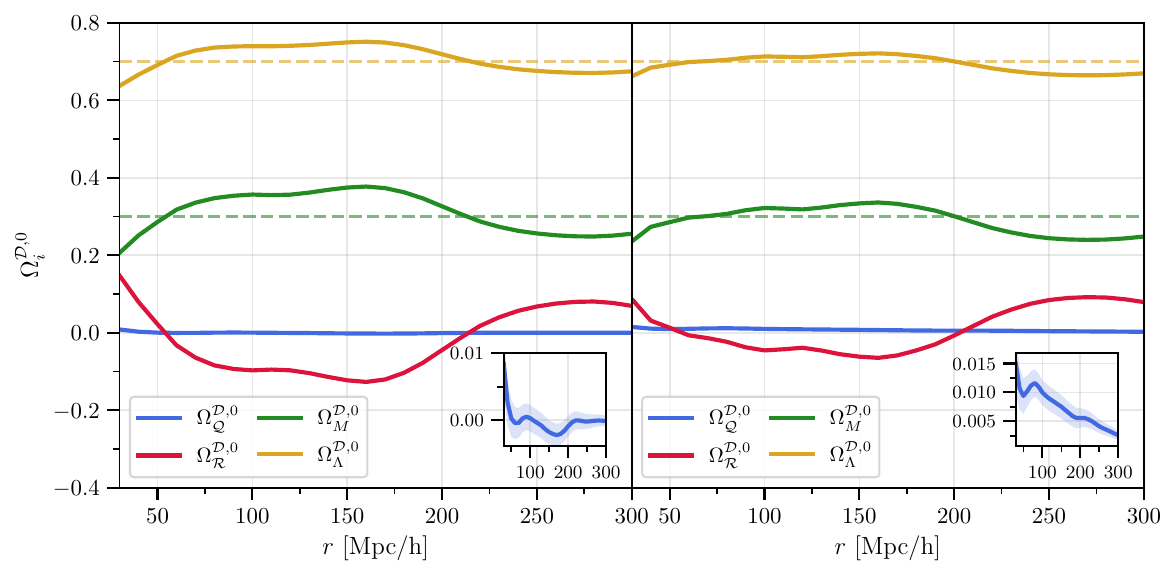}{1.0\textwidth}{}
    \vspace{0.6em}
    \fig{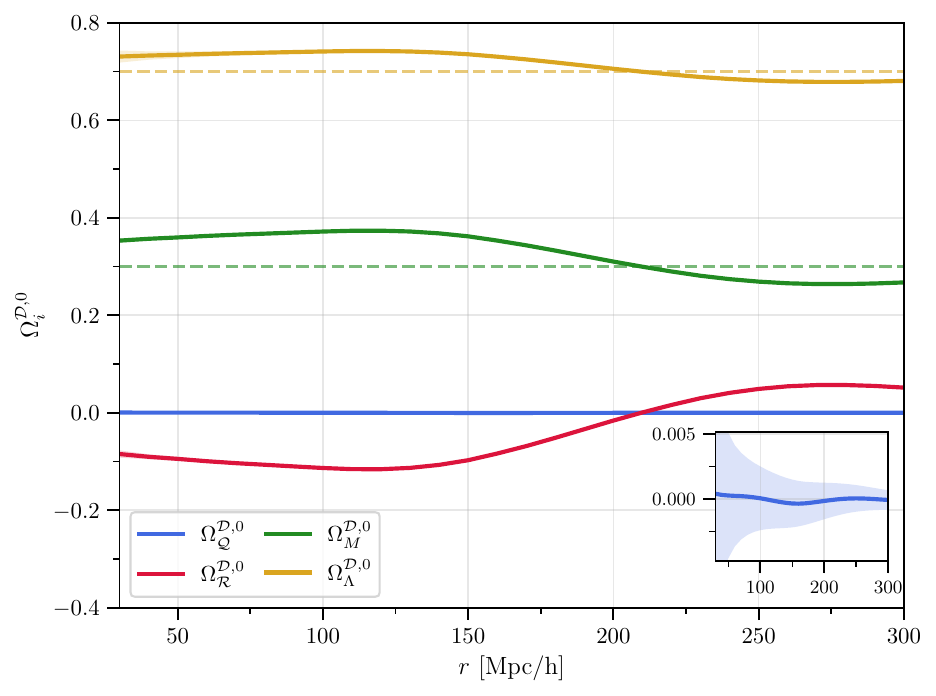}{0.6\textwidth}{}
    \caption{The present-day averaged energy budget, as measured by the $\Omega^{\CD,0}_i$ parameters, and computed within concentric spherical averaging domains $\CD$ of varying comoving radius $r$, from $r = 30$ Mpc/$h$ to $r = 300$ Mpc/$h$, within various robustness tests. The green and yellow horizontal dashed lines in each panel represent the $\Lambda$CDM background values used in the reconstruction for $\Omega_M^0$ and $\Omega_\Lambda^0$, respectively. The zoomed-in insets show $\Omega^{\CD,0}_\mathcal{Q}$, with the shaded areas denoting the $1\sigma$ statistical uncertainty bands. The top row shows the results obtained after applying to the CF4++ reconstruction either quenching corrections on the density contrast (left panel) or the cell-excision procedure (right panel). The cells containing nonlinear density contrasts --- affected by these quenching or excision procedures --- represent about $7 \%$ of the CF4++ within $r = 300$ Mpc/$h$, and about $21 \%$ within $r = 150$ Mpc/$h$. The bottom plot shows the results of employing the $16$-cell sliding-window approach to the CF4++ results. See the main text of Appendix~\ref{app:A} for more details about these three test scenarios.}
    \label{fig:OmegaDi_All}
\end{figure*}

In Fig.~\ref{fig:OmegaDi_All} we show the resulting $\Omega_i^{\mathcal D,0}$ parameters after applying the three controlled regularization procedures above, namely, density quenching (top left), cell excision (top right), and sliding-window smoothing (bottom). The two procedures directly based on the density contrast act primarily as amplitude regularizations. Compared to the baseline analysis, they reduce the amplitude of the fluctuations in $\Omega_i^{\CD,0}$, but they neither change their order of magnitude nor erase the qualitative scale dependence, including the comoving radii at which sign changes in $\Omega_{\Rsp}^{\CD,0}$ occur. By contrast, the $125~\mathrm{Mpc}/h$ sliding-window smoothing constitutes an explicitly nonlocal filtering of the PV and density fields, and correspondingly washes out smaller-scale structures. In particular, the local-void signature below $50~\mathrm{Mpc}/h$ is suppressed within the spatial curvature and averaged density signal as a result. The resulting $\Omega_i^{\CD,0}$ curves become markedly smoother, with overall weakened fluctuations. Nevertheless, these fluctuations still remain of the same order as in the main analysis.

Additionally, the kinematical backreaction contribution, $\Omega_{\mathcal Q}^{\mathcal D,0}$, remains subdominant in all cases (see insets). In particular, the sliding-window smoothing further suppresses the kinematical backreaction amplitude by roughly $1$ order of magnitude. This is expected, since $\Omega_{\mathcal Q}^{\CD,0}$ is sourced by the expansion rate variance and shear of the PV field within the averaging domain $\mathcal D$; both are reduced by the smoothing, which attenuates small- and intermediate-scale gradients in the PV field. Furthermore, we note that the cell-excision strategy (top right) also affects $\Omega_{\mathcal Q}^{\mathcal D,0}$. This remains positive across all scales, rather than fluctuating around zero (as in the other analyses), although it remains small compared to the other contributions to the present-day energy budget.

Accordingly, we conclude that our main qualitative findings in the baseline analysis are robust to these regularization procedures and are not driven by Eulerian linear-theory artifacts. Quantitatively, the precise amplitudes and the transition radii of the features are, as expected, sensitive to the adopted coarse-graining scale --- most markedly for the explicitly nonlocal smoothing at a $125~\mathrm{Mpc}/h$ scale. Qualitatively, however, all three regularizations preserve the same overall structure of the present-day averaged energy budget, including the order-of-magnitude hierarchy among the terms: (i) $\Omega_{\Rsp}^{\CD,0}$ remains a significant, scale-dependent contribution with the same sign-changing behavior (excluding smoothed-out scales); (ii) nonnegligible fluctuations in $\Omega_M^{\CD,0}$ persist; and (iii) $\Omega_{\mathcal Q}^{\CD,0}$ remains a subdominant contribution.

\newpage

\section{Robustness against numerical artifacts}
\label{app:B}

In this appendix, we check the impact of the finite-difference scheme employed within our analysis, given the coarse nature of the reconstruction grid. Specifically, our main results are based on a second-order scheme for the computation of the PV gradients on the reconstruction grid, which are then employed in the calculation of the expansion scalar, $\Theta$, and the shear tensor components, $\sigma_{\mu\nu}$. Here, for comparison, we implement a fourth-order scheme for these gradients and repeat the same analysis described in the main text.

\begin{figure*}[htb!]
    \centering
    \fig{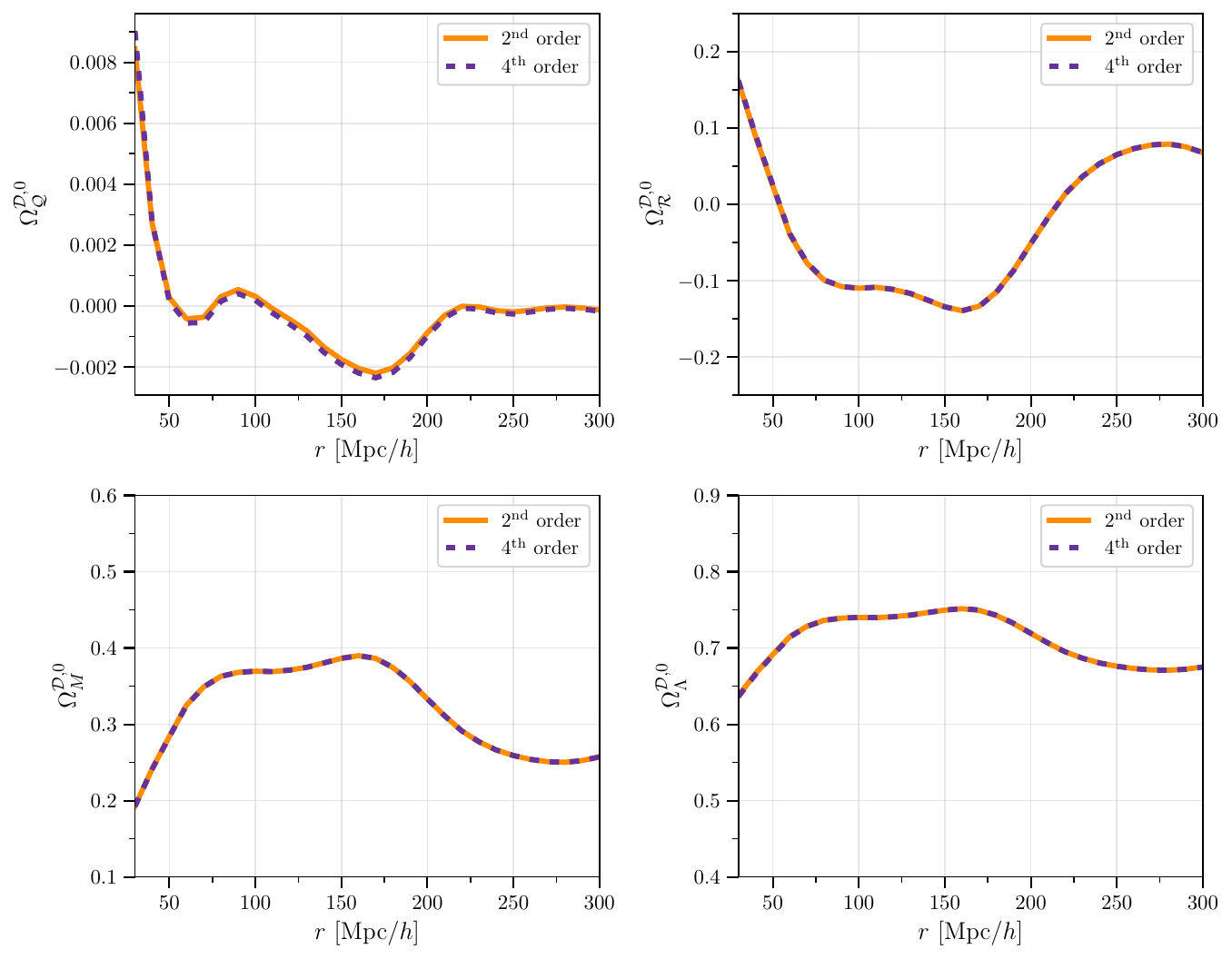}{1.0\textwidth}{}
    \caption{The present-day averaged energy budget inferred  using either a second-order (solid orange lines) or fourth-order (purple dashed lines) finite-difference schemes for computing the PV gradients. Each panel shows a specific $\Omega^{\CD,0}_i$ parameter: $\Omega^{\CD,0}_\CQ$, $\Omega^{\CD,0}_\CR$, $\Omega^{\CD,0}_M$, and $\Omega^{\CD,0}_\Lambda$, respectively, from left to right and top to bottom.}
    \label{fig:Numerics}
\end{figure*}

The results are shown in Fig.~\ref{fig:Numerics}. Differences are found to be negligible and are only barely visible (as relative differences) for $\Omega^{\CD,0}_\CQ$ due to its smaller amplitude and direct dependence on the PV gradients. Overall, we conclude that our results are robust with respect to the numerical scheme implemented.

\bigskip

\end{document}